\documentclass[
 reprint,
superscriptaddress,
 amsmath,amssymb,
 aps,
]{revtex4-2}

\usepackage{graphicx}
\usepackage{dcolumn}
\usepackage{bm}

\begin{document}

\preprint{APS/123-QED}

\title{Isospin breaking in the $^{71}$Kr and $^{71}$Br mirror system}

\author{A. Algora}%
\email{algora@ific.uv.es}
\affiliation{Instituto de Fisica Corpuscular, CSIC-Universitat de Val\'encia, E-46071 Val\'encia, Spain}
\affiliation{%
HUN-REN Institute for Nuclear Research (ATOMKI), H-4001 Debrecen, Hungary}%

\author{A. Vit\'ez-Sveiczer}%
\affiliation{%
HUN-REN Institute for Nuclear Research (ATOMKI), H-4001 Debrecen, Hungary}%
\affiliation{University of Debrecen, PhD school of Physics, H-4026, Debrecen, Hungary} 

\author{A. Poves}%
\affiliation{Departamento de F\'sica Te\'orica and IFT-UAM/CSIC, Universidad Aut\'onoma de Madrid, 28049 Madrid, Spain}%

\author{G. G. Kiss}%
\email{ggkiss@atomki.hu}
\affiliation{%
HUN-REN Institute for Nuclear Research (ATOMKI), H-4001 Debrecen, Hungary}%
\affiliation{%
RIKEN Nishina Center, Wako, Saitama 351-0198, Japan}%

\author{B. Rubio}%
\affiliation{Instituto de Fisica Corpuscular, CSIC-Universitat de Val\'encia, E-46071 Val\'encia, Spain}%

\author{G. de Angelis}%
\affiliation{Istituto Nazionale di Fisica Nucleare, Laboratori Nazionali di Legnaro, Legnaro I-35020, Italy}%

\author{F. Recchia}%
\affiliation{Istituto Nazionale di Fisica Nucleare, Sezione di Padova, Padova I-35131, Italy}%
\affiliation{Dipartimento di Fisica dell Universit\'a degli Studi di Padova, Padova I-35131, Italy}

\author{S. Nishimura}%
\affiliation{RIKEN Nishina Center, Wako, Saitama 351-0198, Japan}%

\author{T. Rodriguez}%
\affiliation{Universidad Complutense de Madrid, 28040, Madrid, Spain}%

\author{P. Sarriguren}%
\affiliation{Instituto de Estructura de la Materia, CSIC, 28006, Madrid, Spain}%

\author{J. Agramunt}%
\affiliation{Instituto de Fisica Corpuscular, CSIC-Universitat de Val\'encia, E-46071 Val\'encia, Spain}%

\author{V. Guadilla}%
\affiliation{Instituto de Fisica Corpuscular, CSIC-Universitat de Val\'encia, E-46071 Val\'encia, Spain}%
\affiliation{Faculty of Physics, University of Warsaw, 02-093, Warsaw, Poland}%

\author{A. Montaner-Piz\'a}%
\affiliation{Instituto de Fisica Corpuscular, CSIC-Universitat de Val\'encia, E-46071 Val\'encia, Spain}%

\author{A. I. Morales}%
\affiliation{Instituto de Fisica Corpuscular, CSIC-Universitat de Val\'encia, E-46071 Val\'encia, Spain}%

\author{S. E. A. Orrigo}%
\affiliation{Instituto de Fisica Corpuscular, CSIC-Universitat de Val\'encia, E-46071 Val\'encia, Spain}%

\author{D. Napoli}%
\affiliation{Istituto Nazionale di Fisica Nucleare, Laboratori Nazionali di Legnaro, Legnaro I-35020, Italy}%

\author{S. M. Lenzi}%
\affiliation{Istituto Nazionale di Fisica Nucleare, Sezione di Padova, Padova I-35131, Italy}%
\affiliation{Dipartimento di Fisica dell Universit\'a degli Studi di Padova, Padova I-35131, Italy}

\author{A. Boso}%
\affiliation{Istituto Nazionale di Fisica Nucleare, Sezione di Padova, Padova I-35131, Italy}%
\affiliation{Dipartimento di Fisica dell Universit\'a degli Studi di Padova, Padova I-35131, Italy}

\author{V. H. Phong}%
\affiliation{RIKEN Nishina Center, Wako, Saitama 351-0198, Japan}%
\affiliation{Faculty of Physics, VNU Hanoi University of Science, 334 Nguyen Trai, Thanh Xuan, Hanoi, Vietnam}

\author{J. Wu}%
\affiliation{RIKEN Nishina Center, Wako, Saitama 351-0198, Japan}%

\author{P.-A. S\"oderstr\"om}%
\affiliation{RIKEN Nishina Center, Wako, Saitama 351-0198, Japan}%

\author{T. Sumikama}%
\affiliation{RIKEN Nishina Center, Wako, Saitama 351-0198, Japan}%

\author{H. Suzuki}%
\affiliation{RIKEN Nishina Center, Wako, Saitama 351-0198, Japan}%

\author{H. Takeda}%
\affiliation{RIKEN Nishina Center, Wako, Saitama 351-0198, Japan}%

\author{D. S. Ahn}%
\affiliation{RIKEN Nishina Center, Wako, Saitama 351-0198, Japan}%

\author{H. Baba}%
\affiliation{RIKEN Nishina Center, Wako, Saitama 351-0198, Japan}%

\author{P. Doornenbal}%
\affiliation{RIKEN Nishina Center, Wako, Saitama 351-0198, Japan}%

\author{N. Fukuda}%
\affiliation{RIKEN Nishina Center, Wako, Saitama 351-0198, Japan}%

\author{N. Inabe}%
\affiliation{RIKEN Nishina Center, Wako, Saitama 351-0198, Japan}%

\author{T. Isobe}%
\affiliation{RIKEN Nishina Center, Wako, Saitama 351-0198, Japan}%

\author{T. Kubo}%
\affiliation{RIKEN Nishina Center, Wako, Saitama 351-0198, Japan}%

\author{S. Kubono}%
\affiliation{RIKEN Nishina Center, Wako, Saitama 351-0198, Japan}%

\author{H. Sakurai}%
\affiliation{RIKEN Nishina Center, Wako, Saitama 351-0198, Japan}%

\author{Y. Shimizu}%
\affiliation{RIKEN Nishina Center, Wako, Saitama 351-0198, Japan}%

\author{S. Chen}%
\affiliation{RIKEN Nishina Center, Wako, Saitama 351-0198, Japan}%

\author{B. Blank}%
\affiliation{Centre d Etudes Nucl\'eaires de Bordeaux-Gradignan, UMR 5797 Universit\'e de Bordeaux CNRS/IN2P3, 19 Chemin du Solarium, CS10120,
33175 Gradignan Cedex, France}%

\author{P. Ascher}%
\affiliation{Centre d Etudes Nucl\'eaires de Bordeaux-Gradignan, UMR 5797 Universit\'e de Bordeaux CNRS/IN2P3, 19 Chemin du Solarium, CS10120,
33175 Gradignan Cedex, France}%

\author{M. Gerbaux}%
\affiliation{Centre d Etudes Nucl\'eaires de Bordeaux-Gradignan, UMR 5797 Universit\'e de Bordeaux CNRS/IN2P3, 19 Chemin du Solarium, CS10120,
33175 Gradignan Cedex, France}%

\author{T. Goigoux}%
\affiliation{Centre d Etudes Nucl\'eaires de Bordeaux-Gradignan, UMR 5797 Universit\'e de Bordeaux CNRS/IN2P3, 19 Chemin du Solarium, CS10120,
33175 Gradignan Cedex, France}%

\author{J. Giovinazzo}%
\affiliation{Centre d Etudes Nucl\'eaires de Bordeaux-Gradignan, UMR 5797 Universit\'e de Bordeaux CNRS/IN2P3, 19 Chemin du Solarium, CS10120,
33175 Gradignan Cedex, France}%

\author{S. Gr\'evy}%
\affiliation{Centre d Etudes Nucl\'eaires de Bordeaux-Gradignan, UMR 5797 Universit\'e de Bordeaux CNRS/IN2P3, 19 Chemin du Solarium, CS10120,
33175 Gradignan Cedex, France}%

\author{T. Kurtuki\'an Nieto}%
\affiliation{Instituto de Estructura de la Materia, CSIC, 28006, Madrid, Spain}%
\affiliation{Centre d Etudes Nucl\'eaires de Bordeaux-Gradignan, UMR 5797 Universit\'e de Bordeaux CNRS/IN2P3, 19 Chemin du Solarium, CS10120,
33175 Gradignan Cedex, France}%

\author{C. Magron}%
\affiliation{Centre d Etudes Nucl\'eaires de Bordeaux-Gradignan, UMR 5797 Universit\'e de Bordeaux CNRS/IN2P3, 19 Chemin du Solarium, CS10120,
33175 Gradignan Cedex, France}%

\author{W. Gelletly}%
\affiliation{Instituto de Fisica Corpuscular, CSIC-Universitat de Val\'encia, E-46071 Val\'encia, Spain}%
\affiliation{Department of Physics, University of Surrey, Guildford GU2 7XH, United Kingdom}%

\author{Zs. Dombrádi}%
\affiliation{HUN-REN Institute for Nuclear Research (ATOMKI), H-4001 Debrecen, Hungary}%

\author{Y. Fujita}%
\affiliation{Research Center for Nuclear Physics, Osaka University, Ibaraki, Osaka 567-0047, Japan}%
\affiliation{Department of Physics, Osaka University, Toyonaka, Osaka 560-0043, Japan}

\author{M. Tanaka}%
\affiliation{Department of Physics, Osaka University, Toyonaka, Osaka 560-0043, Japan}%

\author{P. Aguilera}%
\affiliation{Istituto Nazionale di Fisica Nucleare, Sezione di Padova, Padova I-35131, Italy}%
\affiliation{Comisi\'on Chilena de Energ\'ia Nuclear, Casilla 188-D, Amun\'ategui 95, Santiago Centro, Santiago, Chile}%

\author{F. Molina}%
\affiliation{Comisi\'on Chilena de Energ\'ia Nuclear, Casilla 188-D, Amun\'ategui 95, Santiago Centro, Santiago, Chile}%

\author{J. Eberth}%
\affiliation{Institute of Nuclear Physics, University of Cologne, D-50937 Cologne, Germany}%

\author{F. Diel}%
\affiliation{Institute of Nuclear Physics, University of Cologne, D-50937 Cologne, Germany}%

\author{D. Lubos}%
\affiliation{Physik Department E12, Technische Universit\"at M\"unchen, D-85748 Garching, Germany}%

\author{C. Borcea}%
\affiliation{National Institute for Physics and Nuclear Engineering IFIN-HH, P.O. Box MG-6, Bucharest-Magurele, Romania}%

\author{E. Ganioglu}%
\affiliation{Department of Physics, Istanbul University, Istanbul 34134, Turkey}%

\author{D. Nishimura}%
\affiliation{Department of Natural Sciences, Tokyo City University of Science, Setagaya-ki, Tokyo 158-8557, Japan}%

\author{H. Oikawa}%
\affiliation{Department of Physics, Tokyo University of Science, Noda, Chiba 278-8510, Japan}%

\author{Y. Takei}%
\affiliation{Department of Physics, Tokyo University of Science, Noda, Chiba 278-8510, Japan}%

\author{S. Yagi}%
\affiliation{Department of Physics, Tokyo University of Science, Noda, Chiba 278-8510, Japan}%

\author{W. Korten}%
\affiliation{CEA Saclay, IRFU, SPhN, 91191 Gif-sur-Yvette, France}%

\author{G. de France}%
\affiliation{Grand Acc\'el\'erateur National d Ions Lourds, B.P. 55027, F-14076 Caen Cedex 05, France}%

\author{P. Davies}%
\affiliation{Department of Physics, University of York, Heslington, York YO10 5DD, United Kingdom}%

\author{J. Liu}%
\affiliation{Department of Physics, University of Hong Kong, Pokfulam, Hong Kong}%

\author{J. Lee}%
\affiliation{Department of Physics, University of Hong Kong, Pokfulam, Hong Kong}%

\author{T. Lokotko}%
\affiliation{Department of Physics, University of Hong Kong, Pokfulam, Hong Kong}%

\author{I. Kojouharov}%
\affiliation{GSI, Planckstrasse 1, D-64291 Darmstadt, Germany}%

\author{N. Kurz}%
\affiliation{GSI, Planckstrasse 1, D-64291 Darmstadt, Germany}%

\author{H. Schaffner}%
\affiliation{GSI, Planckstrasse 1, D-64291 Darmstadt, Germany}%

\author{A. Kruppa}%
\affiliation{%
HUN-REN Institute for Nuclear Research (ATOMKI), H-4001 Debrecen, Hungary}%
\affiliation{%
RIKEN Nishina Center, Wako, Saitama 351-0198, Japan}%







\date{\today}

\begin{abstract}
Isospin symmetry is a fundamental concept in nuclear physics. 
Even though isospin symmetry is partially broken, it holds approximately for most nuclear systems, which makes exceptions very interesting from the nuclear structure perspective.  In this framework, it is expected that the spins and parities of the ground states of mirror nuclei should be the same, in particular for the simplest systems where a proton is exchanged with a neutron or vice versa. In this work, we present evidence that this assumption is broken in the mirror pair $^{71}$Br and $^{71}$Kr system. Our conclusions are based on a high-statistics $\beta$ decay study of $^{71}$Kr and on state-of-the-art shell model calculations. In our work, we also found evidence of a new state in $^{70}$Se, populated in the $\beta$-delayed proton emission process which can be interpreted as the long sought coexisting 0$^+$ state. 
\end{abstract}

\maketitle


~
\newpage
~
\newpage
~

Since its introduction in 1937 \cite{Wigner}, isospin as a quantum number and the associated isospin symmetry have played a key role in our understanding of particle and nuclear physics. In the nuclear physics realm, the symmetry is expected to be only approximate, partially broken due to the presence of the Coulomb interaction, isospin breaking components of the strong force and the mass difference between the proton and neutron. Nevertheless, the breaking of the symmetry is limited. 
Accordingly, 
the interpretation of a large variety of phenomena in nuclei can be based on the conservation of T, the isospin quantum number. 

Isobaric systems in which the proton and neutron numbers are interchanged, the so-called mirror nuclei, are of particular interest for the study of isospin symmetry (see \cite{bl,zlmp} and references therein). For example, it is expected that the structures of their ground states should be very similar and this has been observed to be the general rule so far.  Among all  mirror systems studied, there are only two exceptional cases that have been identified where the isospin symmetry breaks in this particular sense. One case is the T=1 $^{16}$F-$^{16}$N mirror pair, a special system where $^{16}$F is particle unbound and $^{16}$N is particle bound \cite{Stefan}. The symmetry breaking in this case is explained as a consequence of the Coulomb interaction, by an effect known as the Thomas-Ehrmann shift \cite{Thomas1,Ehrman,Thomas2}, related to the more loosely bound proton in $^{16}$F. 
In those mirror nuclei the l=0 and l=2 states are close to degeneracy. Since these orbitals have different radial extensions, their Coulomb displacement energies are different thus modifying the level ordering.
The second and until now the only known example in which both nuclear systems are bound is the recently studied $^{73}$Br/$^{73}$Sr mirror pair with T=3/2 \cite{hoff}. This last case was identified indirectly by studying the proton emission of states in $^{73}$Rb that indicate the assignment of $J^\pi$ =(5/2$^-$) to the ground state of $^{73}$Sr, to be compared with the $J^\pi$=1/2$^-$ assignment to $^{73}$Br in its ground state. These cases are extraordinary and pinpoint specific nuclear structure effects. 


In this letter we present a detailed $\beta$ decay study of the close to drip line nucleus $^{71}$Kr. Our work shows that the experimental data can only be interpreted theoretically if $^{71}$Kr has a $J^\pi$ assignment in its ground state that differs from its mirror $^{71}$Br. The evidence is based on a careful spectroscopic study and on the comparison with state-of-the-art shell model calculations. This represents the first example of established mirror symmetry breaking in pairs with T=1/2. Considering that a T=1/2 system is the case where only one nucleon is interchanged, one might argue that seeing isospin effects in such systems is less probable, since it represents the simplest possible change. Mirror decay studies with T=1/2 are also relevant for testing the Conserved Vector Current Hypothesis (CVC) and the determination of the $|V_{ud}|$ matrix element \cite{Naviliat,Severijns,HardyCVC}. Our work shows that the A=71 system presents the lowest confirmed ground state to ground state branching ratio in mirror T=1/2  nuclei \cite{Severijns}. The high-quality decay data obtained in this study also allowed us to identify new states populated in the $\beta$-delayed proton emission of $^{71}$Kr. One of those states, can be interpreted as a long sought coexisting 0$^+$ state in $^{70}$Se \cite{Smallcombe}.


The $\beta$ decay of $^{71}$Kr represents a challenging case for study. $^{71}$Kr is a neutron deficient nucleus, lying in the vicinity of the N=Z line, which is not easy to produce in high abundance. 
The dominant decay channel of $^{71}$Kr is $\beta$ decay, that should proceed mainly by a transition from its ground state to the ground state of the mirror $^{71}$Br based on the grounds of isospin symmetry. Previous  related works have not been free from controversy. The first detailed $\beta$ decay study of $^{71}$Kr was performed at ISOLDE by Oinonen {\it et al.} \cite{Oinonen}. In this work the spin-parity of the ground state of $^{71}$Kr was assigned 5/2$^-$ based on the assumption that the ground state of $^{71}$Kr should have the same character as that of $^{71}$Br, previously known from in-beam studies. Mirror decays are characterized by a dominant branching ratio (BR) to the isobaric analogue state of the order of 90 $\%$ or higher \cite{Severijns}, and the ground state to ground state branch deduced in Oinonen {\it et al.}, was surprisingly small (82.1(16) $\%$) \cite{Oinonen}. This triggered an alternative interpretation by Urkedal and Hamamoto \cite{Urkedal}, who proposed that there might be a spin inversion in the mirror pair due to strong nuclear shape effects occurring in the region.  This last assignment was reviewed later in the work of Fischer {\it et al.} \cite{Fischer}, who performed a careful in-beam study of $^{71}$Br and extrapolated the knowledge obtained to the low-lying states in $^{71}$Kr, assigning back 5/2$^-$ to the ground state, and questioning the interpretation of Urkedal and Hamamoto. This assignment has been retained in a recent work by S. Waniganeththi {\it et al.} \cite{Waniganeththi}. 



$^{71}$Kr was produced at the RIKEN Nishina Center in the fragmentation of a  $^{78}$Kr beam (E$_{beam}$ = 345 MeV) impinging on a 5-mm-thick Be target. 
During the experiment typical beam intensities were of the order of I$_{beam}$ $\sim$ 40 pnA. The reaction products were separated and selected in the BigRIPS separator \cite{fukuda13} using the $\Delta$E-B$\rho$-ToF method, based on an event-by-event measurement of the energy loss ($\Delta$E), magnetic rigidity (B$\rho$), and time of flight (ToF) of the ions. About 9.85 million $^{71}$Kr ions were implanted into the WAS3ABi array \cite{was3abi}, consisting of three double-sided silicon strip detectors (DSSSD). Implant events were correlated with identified ions in BigRIPS. Each DSSSD had a thickness of 1 mm with an active area segmented into 60 horizontal and 40 vertical strips. The DSSSDs were also used to detect $\beta$ particles and $\beta$-delayed protons.


\begin{figure*}[htb]
\resizebox{2\columnwidth}{!}{\rotatebox{0}{\includegraphics[clip=]{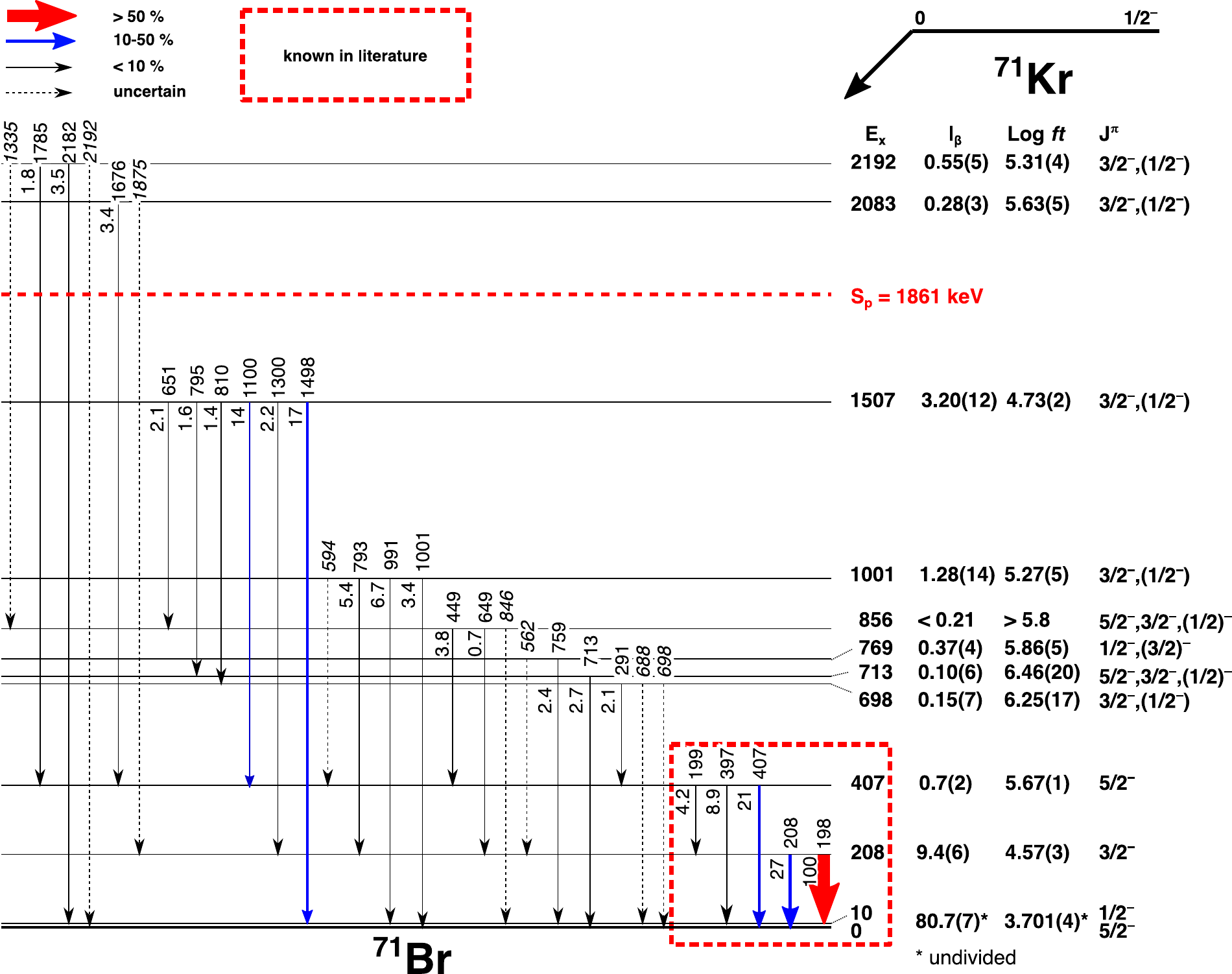}}}
\caption{ 
\label{fig:decayscheme}
Partial level-scheme of $^{71}$Br derived from the $\beta$ decay of $^{71}$Kr. Excitation energies, $\beta$ feedings, Logfts and spin-parity assignments of the observed states are indicated on the right-hand side. The widths of the arrows are proportional to the absolute intensities of the $\gamma$-rays. The region framed by the red dashed line indicates the limited nuclear structure information available before our experiment. 
}
\end{figure*}  

The $\beta$ decay of $^{71}$Kr populates states in $^{71}$Br and in $^{70}$Se through the $\beta$-delayed proton emission process. The $\gamma$ rays emitted following the $\beta$ decays were detected using the Euroball-RIKEN Cluster Array (EURICA) spectrometer \cite{eurica} which consisted of 84 HPGe crystals arranged in twelve clusters at a nominal distance of 22 cm from the centre of the DSSSDs. The absolute $\gamma$ detection efficiency was derived using calibrated $^{60}$Co, $^{133}$Ba and $^{152}$Eu sources and found to be about 8\% at 1332 keV in add-back mode. 

Implantation events in WAS3ABi were correlated in time with  $\beta$ decay events ($\beta$ particle or $\beta$-delayed proton) taking place in the same x and y strips of the DSSSDs where the implantation event was observed. The maximum time difference between an implantation and the corresponding detection of a $\beta$ particle or proton was fixed at 20 s and the $\beta$-delayed $\gamma$ events were recorded up to 800 ns after the decay.

Three methods, resulting in a consistent value, were used to determine the half-life of $^{71}$Kr. Namely, it was determined from the time distribution of implantation-$\beta$ (i-$\beta$) correlations fitted with the Bateman-formula and a background term, extracted from a linear fit to the backward-time distribution of i-$\beta$ correlations. The $\beta$ particles and the $\beta$-delayed protons were distinguished by establishing an energy cut at E$_{cut}$ = 1400keV on the WAS3ABI signals. Events with energies lower than this threshold were identified as $\beta$ particles and the rest as $\beta$-delayed protons. The cut energy was established by comparing the $\beta$-decay spectrum with systems with similar Q value and no $\beta$-delayed proton emission. The $t_{1/2}$ value was also extracted from the decay-curves of 11 intense $\beta$-delayed $\gamma$ transitions associated with $^{71}$Br. Finally, the time distribution of implantation-proton (i-p) correlations was fitted with a exponential function and a background term, determined from the backward-time distribution of i-p correlations. For more details on these procedures see \cite{71Kr_halflife,Orrigo1}. Using this fit the $\beta$-delayed proton branch ($\epsilon_p$) value was derived also and found to be $\epsilon_p=2.77(24)$\%. This value is consistent with the recent results by Waniganeththi {\it et al.} \cite{Waniganeththi}.
The derived more precise $t_{1/2}$ values are  $t_{1/2}^{i\beta} = 94.50 \pm 0.06~(\text{stat.})~~^{+0.21}_{-0.23}~(\text{sys.})$ ms, $t_{1/2}^{i\beta\gamma} = 94.7 \pm 0.5~(\text{stat.})~~^{+1.6}_{-1.5}~(\text{sys.})$ ms and $t_{1/2}^{ip} = 94.08 \pm 0.22~(\text{stat.})~\pm0.33~(\text{sys.})$ ms deduced from fitting the i-$\beta$, i-$\beta$-$\gamma$ and i-p time correlations, respectively (see Supplemental Material Fig. SM.I for more details). 

Prior to our experiment, a limited amount of information was available on the $\beta$ decay of $^{71}$Kr. 
Only five $\gamma$ transitions, connecting the three, low-lying excited states and the ground state were known \cite{Waniganeththi,Oinonen}. In our study $\gamma$ rays corresponding to the de-excitation of states populated in the 
$\beta$ decay were identified primarily using $\gamma\gamma$ coincidences with the known transitions. Then, direct transitions connecting the newly found excited states with the ground state were sought. Figure SM.II 
in Supplemental Material \cite{SupMat} shows the measured $\gamma$ spectrum correlated with $^{71}$Kr decays. Altogether 25 new $\gamma$ transitions, connecting 8 excited states, were identified for the first time. 
$\gamma$-gated spectra were used to measure and verify branching ratios of transitions de-exciting the same levels. The deduced level scheme is shown in Fig.\ref{fig:decayscheme}. The total number of $\beta$ particles emitted in the decay of the $^{71}$Kr was derived from  the integration of the appropriate component of the i-$\beta$ correlations in the fit using the Bateman equation. The $I_\beta$, log$ft$ and B(GT) values (see Fig. \ref{fig:decayscheme} and Supplemental Material \cite{SupMat}), were calculated using the absolute $\gamma$-ray intensities, $t_{1/2}=94.50(24)$ ms and $Q_{EC}= 10~180(130)$ keV \cite{Wang}. The contribution of the super-allowed Fermi-transition to the ground state doublet was calculated using the Fermi sum rule \cite{Osterfeld_1992}, assuming that it was exhausted (B(F)=1).

\begin{figure}[htb]
\resizebox{1.0\columnwidth}{!}{\rotatebox{0}{\includegraphics[clip=]{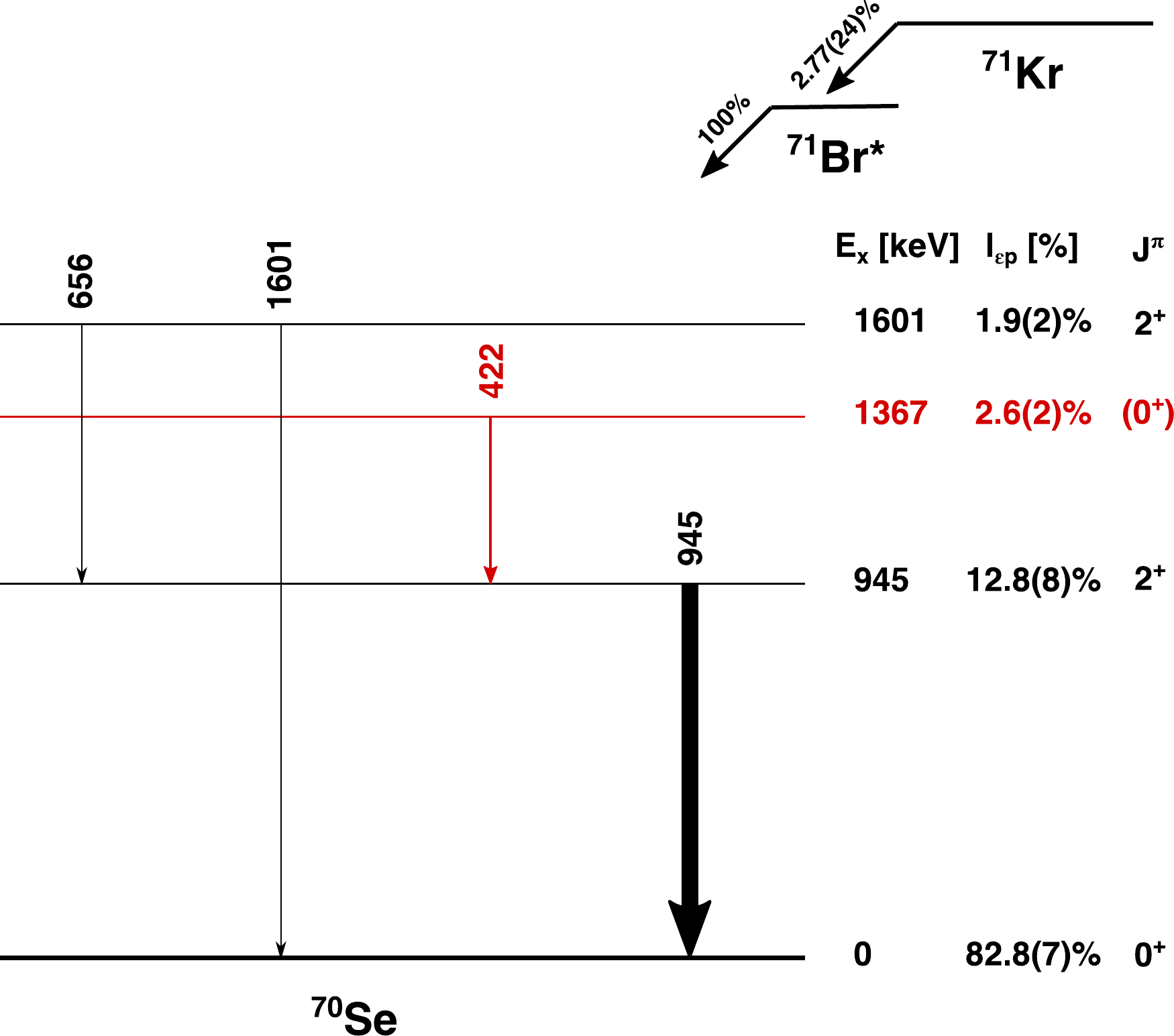}}}
\caption{ 
\label{fig:p_branch}
Level-scheme of $^{70}$Se populated in the beta delayed proton decay of $^{71}$Kr. Excitation energies, relative feedings ($\epsilon_p \equiv 100\%$) and spins and parities ($J^{\pi}$) of the observed states are indicated on the right-hand side of the levels. The arrow widths are proportional to absolute intensities of the $\gamma$ rays. The absolute $\beta$ delayed proton branch is $\epsilon_p$= 2.77(24) \% (see the text for details).}
\end{figure}

Gamma rays following the $\beta$-delayed proton emission were measured, too. Three known transitions (at energies of E$_{\gamma}$=945 keV, 656 keV and 1601 keV) in the daughter nucleus $^{70}$Se were observed, as well as a previously unknown transition at $E_\gamma = 422$ keV. $\gamma\gamma$ coincidences and absolute intensities were  used to deduce the decay-scheme proposed in Fig. \ref{fig:p_branch}.


 To interpret the $\beta$ decay data, we have carried out large scale shell model calculations (SM-CI) using the interaction PFSDG-U \cite{pfsdg-u} 
 in the valence space encompassed by the orbits 1p$_{3/2}$,  1p$_{1/2}$,  0f$_{5/2}$,  0g$_{9/2}$, and  1d$_{5/2}$.
 Coulomb and other charge symmetry breaking (CSB) terms were evaluated using the approach presented in \cite{bl,Bentley}. Recently the same framework has been used to explain the inversion of the ground state spin in the T=3/2 A=73 mirrors, from
 $1/2^-$ in  $^{73}$Br to  $5/2^-$ in  $^{73}$Sr in \cite{a73}. 
 
\begin{figure}[htb]
\resizebox{1.0\columnwidth}{!}{\rotatebox{0}{\includegraphics[clip=]{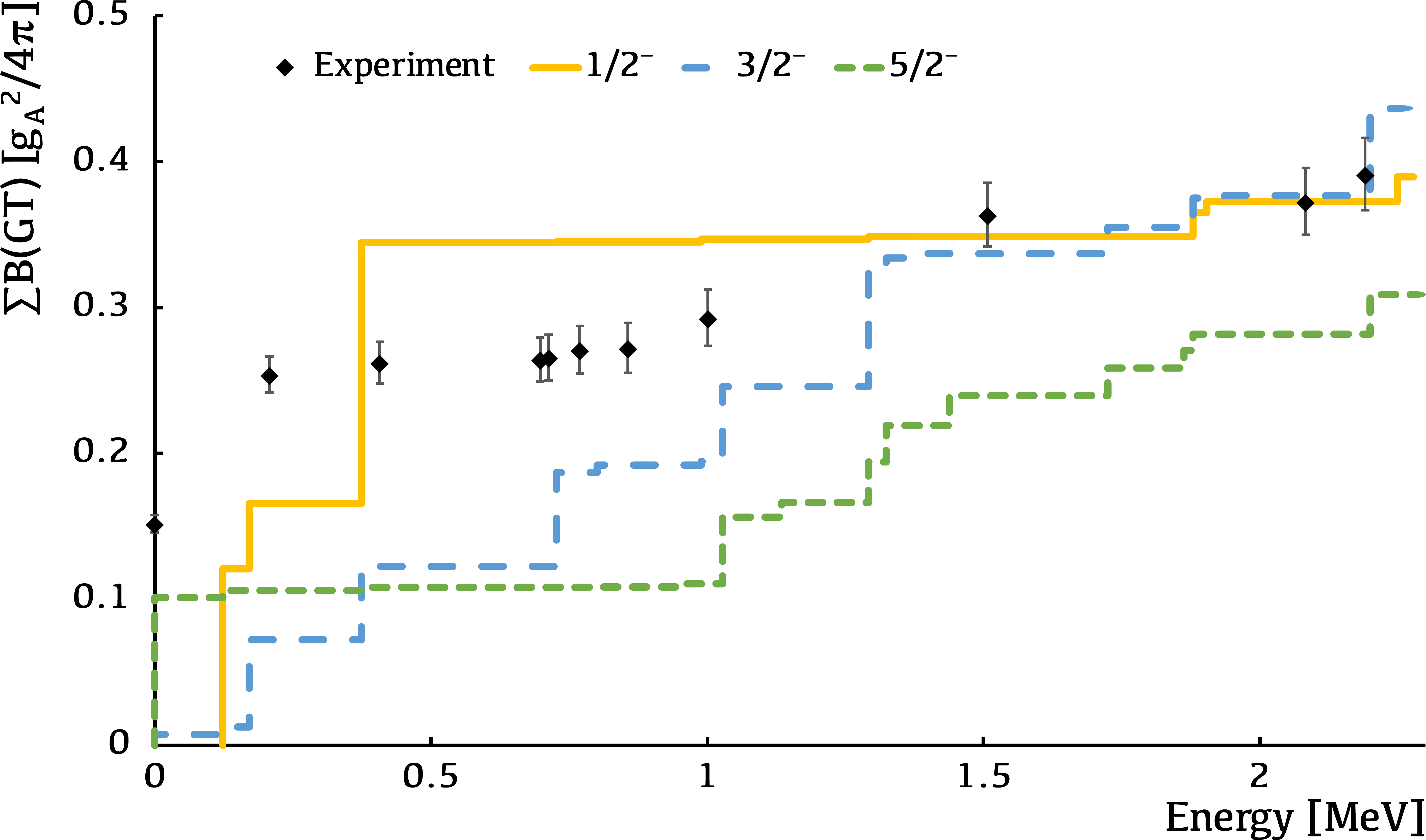}}}
\caption{ 
\label{fig:BGT}
Accumulated GT strength distribution of the $^{71}$Kr $\beta$ decay as a function of the excitation energy in the daughter $^{71}$Br  nucleus. Black symbols indicate the experimental B(GT) distribution. The calculated GT strength distributions correspond to results assuming 5/2$^-$ (green dotted), 3/2$^-$ (blue dashed) and 1/2$^-$ (yellow solid) as ground states in $^{71}$Kr, respectively.}
\end{figure}

As the CSB contributions are fully perturbative in the wave functions, we have ignored them in the calculation of the Gamow-Teller (GT) strength function for the $\beta$ decay of $^{71}$Kr. Given that in the experimental spectrum of $^{71}$Br the  $1/2^-$ is just 10~keV above the $5/2^-$ ground state, we have computed the  GT strength functions for these two states as ground states in $^{71}$Kr. These results are compared with the values deduced from the present experiment in Fig. \ref{fig:BGT}. For completeness,  the $3/2^-$ state was also considered as a possible ground state in $^{71}$Kr, since in $^{71}$Br it is the second excited state at 208~keV. 
 
 
\begin{table*}[tbh]
\begin{center}
  \caption{Calculated energy levels with the PFSDG-U interaction (A=71 column) and considering in successive approximations the electromagnetic and nuclear charge symmetry breaking (CSB) terms (see reference \cite{a73} for more details): (a) Nuclear plus Coulomb two body plus the electromagnetic 
  $\vec{l} \cdot \vec{l}$ and  $\vec{l} \cdot \vec{s}$ contributions to the single particle energies of protons and neutrons; (b)=(a) plus the nuclear CSB
  correction; (c)=(b) plus the radial correction to the mirror energy differences (MED) assigned to  $^{71}$Kr. All energies are given in keV. \label{spectra}}
   \begin{tabular*}{\linewidth}{@{\extracolsep{\fill}}|c|ccccccc|}\hline
  J$^{\pi}$  &  A=71 & $^{71}$Kr (a) & $^{71}$Kr (b) & $^{71}$Kr (c) &  $^{71}$Br (a) & $^{71}$Br (b) & $^{71}$Br (exp) \\ 
\hline
  $5/2^-$  & 0   & 0 & 0  &0  & 0  & 0   & 0   \\
  $1/2^-$   & 171  & 45 &  50 &  23&  33 &  26 &   10 \\
  $3/2^-$  &123  & 131 &  136 & 137 & 182   & 189  & 208  \\
  $9/2^+$   & 1047 & 975 &  998  & 902 & 882  & 877  &  759 \\
\hline
    \end{tabular*}
 \end{center}
 \label{tab:theory}
\end{table*}


As seen in Fig. \ref{fig:BGT}, the comparison of the deduced accumulated $\beta$ strength with state-of-the-art shell model calculations shows a clear preference for the assignment of $J^{\pi}$=$1/2^-$ to the ground state of $^{71}$Kr within the resolution scale of the SM-CI. Only the accumulated $\beta$ strength of this state is able to reproduce the experimental strength pattern. The $\beta$ decay of the $1/2^-$ state shows a dominant Gamow-Teller transition at around 0.3 MeV excitation energy in the daughter nucleus, that remains almost constant up to the Q value. On the contrary the $\beta$ decay of the $5/2^-$ state shows a quite different pattern, with a continuous increase of the strength between 1 and 2 MeV that does not match the experiment. This is a clear indication of the necessity to assign a different spin-parity to the ground state of $^{71}$Kr compared with $^{71}$Br and an evidence of isospin breaking in the mirror system with A=71. Experimentally, this assignment is also supported by the fact, that our higher sensitivity experiment does not show feeding to the known low-lying $7/2^-$ states in $^{71}$Br as it was expected in the case of a $5/2^-$ ground state assignment in $^{71}$Kr \cite{Fischer}. 
In table \ref{tab:theory} I the lowest calculated states in $^{71}$Kr and $^{71}$Br are also provided, showing the impact of the Coulomb and other symmetry breaking terms in the calculated energies in $^{71}$Kr. 
Even though in column (c) the $5/2^-$ state remains the lowest state for $^{71}$Kr after the corrections, its energy separation from the preferred $1/2^-$ state is only 23 keV. The obtained results are consistent with the spin-parity assignment of $1/2^-$ to the ground state suggested in this work within the precision of the shell model calculations.  

To justify the claim of the $J^{\pi}$ assignment based on the comparison with the accumulated GT strength one has to address the question of whether our experimental decay data might suffer from the Pandemonium effect \cite{Hardy} or not. The use of the highly efficient EURICA setup is a good 
choice in this respect, 
but it does not fully guarantee that some $\gamma$ rays might remain undetected (see for example \cite{AlgoraHo} where a highly efficient array was also used). In the present study, we have determined a feeding to the isobaric analogue state of 80.7(7)$\%$, which represents the lowest experimentally confirmed feeding among all studied mirror decays (see Severijns {\it et al.} \cite{Severijns}). But even though it is the smallest among the mirror nuclei, experimentally a $\beta$ feeding of 80.7(7)$\%$ can be considered large.
Our previous experience from total absorption measurements shows that high-resolution data from decays that are dominated by transitions to ground (or low-excitation) states in the daughter nucleus are less likely to suffer from the Pandemonium effect,  see for example $^{102}$Tc, $^{101}$Nb \cite{Algora_lett,Jordan} $^{100}$Tc \cite{Guadilla100}, and $^{96gs}$Y \cite{Guadilla96Y} decays. This can be easily understood, since the higher the fraction of ground state to ground (or low excitation) state feeding of the total, the lower the amount of $\beta$ feeding associated with high excitation levels, reducing the possibility of missing feeding at high excitation in the daughter nucleus. In addition, in the present experiment we can expect that levels populated above the proton separation energy in the daughter nucleus will predominantly decay by $\beta$-delayed proton emission, after a transitional excitation region where competition between proton and $\gamma$ emission can occur (see for example \cite{Dossat,Orrigo2}). Since in $^{71}$Br  the proton separation energy is at a relatively low energy ($S_p$=1861 keV), and the $\beta$ delayed proton branch is relatively small (2.77(24)\%), we can assume that most of the remaining $\beta$ feeding will occur in an excitation energy interval in the daughter, where the efficiency for $\gamma$ detection is high in the EURICA setup leaving little margin for the Pandemonium effect. This provides confidence in the present assignment of the spin-parity of $J^\pi$=1/2$^-$ to the ground state of $^{71}$Kr based on a comparison of the deduced experimental $\beta$ strength with theoretical calculations.

We can now return to the question raised by Urkedal and Hamamoto \cite{Urkedal} related to the possible deformation of $^{71}$Kr in its ground state. 
In the mean field approach based on the Skyrme SLy4 interaction, several minima in the energy-deformation
surface are obtained in both the prolate and oblate sectors. The ground state corresponds to an oblate minimum
consistent with the 1/2$^-$ assignment with a deformation $\beta$=-0.18. On the prolate side the various local minima
correspond to a 3/2$^-$ state with $\beta$=0.14 and to close-lying  excited states 5/2$^-$ and 1/2$^-$ with deformation $\beta$=0.09.


We should also discuss qualitatively the states populated in the $\beta$ delayed proton emission process (see Fig. \ref{fig:p_branch}). The oblate character of the 0$_{1}^+$ and 2$_{1}^+$ states in $^{70}$Se is considered settled experimentally \cite{Ljungvall,Henderson,Fisher2,Smallcombe} in agreement with several model calculations \cite{Bertsch,Hinohara1,Hinohara2}. In the present description we rely on beyond mean field calculations  \cite{Rodriguez}. 
In this framework  the first 0$_{1}^+$ and 2$_{1}^+$ states have a dominant oblate character in agreement with previous studies.  This could explain their strong population in the $\beta$ delayed proton emission process based on the fact that transitions between states with similar deformation should be preferred in the decay. In  the beyond mean field calculations, two additional 0$^+$ states are predicted at 1000 keV and 1400 keV respectively lying at lower excitation energies than the next excited 2$_{2}^+$ and 2$_{3}^+$ states. 
Experimentally we identify the new state at 1367 keV as the 0$_{2}^+$ state in the beyond mean field approach. In the calculations, this last state has large structural overlap with the 0$_{1}^+$ ground state in $^{70}$Se, which can explain the direct $\beta$ delayed proton branch. Based on this we identify the 1367 keV state as the long sought coexisting 0$^+$ state in $^{70}$Se not seen until now \cite{Smallcombe}. According to Fig. SM.III 
in the Supplemental Material \cite{SupMat}, this state has a clear triaxial structure. The experimental assignment of (0$_{2}^+$) is supported on the fact that there is no $\gamma$ transition from this state to the ground state in $^{70}$Se. 

In this letter we have reported on new evidence on mirror-symmetry breaking in nuclei near the proton drip line. 
We have found that the experimentally extracted Gamow-Teller strength in the $\beta$ decay of $^{71}$Kr can only be explained by state-of-the-art shell model calculations if there is an inversion of the two lowest energy states in the T=1/2 A=71 mirror system. 
This case is the first known mirror symmetry breaking example that belongs to the simplest possible category, where just one proton is exchanged with a neutron. The level inversion can be explained quantitatively with shell model calculation including Coulomb and other isospin breaking terms. These conclusions are based on solid experimental evidence obtained from a high-statistics study performed at the RIKEN Nishina Center and state-of-the-art theoretical calculations. The incomplete decay data and limited statistics from previous studies \cite{Oinonen,Waniganeththi} impeded reaching the conclusions achieved here. 



The observed $\beta$ branch to the isobaric analogue state presents the smallest experimentally confirmed $\beta$ branch among all studied T=1/2 mirror nuclei \cite{Severijns}. 
The high statistics $\beta$ delayed proton data points to the existence of a new state in $^{70}$Se, identified here as a long-sought shape coexisting 0$^+_2$ state. This last observation suggests that one might use $\beta$-delayed processes to seek coexisting states that are only fed directly in a very limited way in nuclear reactions. 
The newly identified (0$^+$) state does not solve the puzzle associated with the work of Wimmer {\it et al.} \cite{Wimmer} (see also Lenzi {\it et al.} \cite{Lenzi}). The identified state lies above the 2$^+_1$ state in $^{70}$Se, so it can not affect dramatically the conclusions reached in \cite{Wimmer}.

\begin{acknowledgments}
This work was carried out at the RIBF operated by RIKEN Nishina Center and CNS, University of Tokyo.
We acknowledge the EUROBALL Owners Committee for the loan of germanium detectors and the PreSpec Collaboration for the readout electronics of the cluster detectors. This work was supported by the Spanish MICINN grants FPA2014-52823-C2-1-P, FPA2017-83946-C2-1-P (MCIU/AEI/FEDER); Ministerio de Ciencia e Innovacion grants PID2019-104714GB-C21, PID2022-138297NB-C21; Centro de Excelencia Severo Ochoa CEX2023-001292-S grant funded by MCIU/AEI; $Junta~para~la~Ampliaci\acute{o}n~de~Estudios$ Programme (CSIC JAE-Doc contract) co-financed by FSE, grant CEX2020-001007-S  funded by MCIN/AEI/10.13039/501100011033 and PID2021-127890NB-I00, by NKFIH (K147010), JSPS KAKENHI of Japan (Grant No. 25247045, 20H05648, 22H04946), the STFC (UK) through Grant No. ST/P005314/, the PROMETEO 2019/007 and CIPROM/2022/9 projects. This work was supported by Generalitat Valenciana (CISEJI/2022/25), Ministerio de Ciencia, Innovación y Universidades (CAS22/00114 and CNS2023-144871). 
A. A.  acknowledges partial support of the JSPS Invitational Fellowships for Research in Japan (ID: L1955)
P. S. acknowledges support from MCI/AEI/FEDER,UE (Spain) under grant PID2022-136992NB-I00.. Enlightening discussions with N. Severijns, M. Gonzalez-Alonso, and J. L. Tain are also acknowledged. 
\end{acknowledgments}


\bibliography{references}

\renewcommand\thefigure{SM.\Roman{figure}}    
\setcounter{figure}{0}  

\renewcommand\thetable{SM.\Roman{table}}    
\setcounter{table}{0} 

{\bf Supplemental Material}

\begin{table*}[ht!]
\caption{Table of excited levels and their identified $\gamma$ transitions. Transitions confirmed by $\gamma\gamma$ coincidences are marked in the last coloumn.}
\begin{tabular}{|l|l|l|l|l|l|l|l|l|l|}
\hline
$E_x$ {[}keV{]} & $J^\pi$            & $I_\beta$       & log $ft$          & $B(GT) [G_A^2/4\pi]$ & $B(F)  [G_V^2/4\pi]$ & $E_f$ {[}keV{]} & $E_\gamma$ {[}keV{]} & $I_\text{rel.}$ & Coincidence \\ \hline \hline
0 + 10 (Fermi)  & 1/2–, (5/2-)       & 65.5            & 3.791             &                      & $|N-Z|=1$            &                 &                      &                 &             \\
0 + 10 (GT)     & 5/2-, (1/2)-       & 15.2(7) & 4.41(2) & 0.148(7)             &                      &                 &                      &                 &             \\ \hline
208             & 3/2-               & 9.4(6)          & 4.57(3)           & 0.102(6)             &                      & 10              & 198                  & 100(7)          & C           \\
                &                    &                 &                   &                      &                      & 0               & 208                  & 28.5(19)        & C           \\ \hline
407             & 5/2-               & 0.70(18)        & 5.67(10)          & 0.008(2)             &                      & 208             & 199                  & 4.2(13)         & C           \\
                &                    &                 &                   &                      &                      & 10              & 397                  & 8.9(6)          & C           \\
                &                    &                 &                   &                      &                      & 0               & 407                  & 20.5(11)        & C           \\ \hline
698             & 3/2-, (1/2)-       & \textless{}0.2  & \textgreater{}6.1 & 0.0021(8)            &                      & 407             & 291                  & 2.1(3)          & C           \\
                &                    &                 &                   &                      &                      & 10              & 688                  & \textless 1.1   &            \\
                &                    &                 &                   &                      &                      & 0               & 698                  & 1.1(7)          & C           \\ \hline
713             & 5/2-, 3/2-, (1/2)- & \textless{}0.2  & \textgreater{}6.3 & 0.0013(6)            &                      & 0               & 713                  & 2.7(3)          & C           \\ \hline
769             & 1/2-, 3/2-         & 0.37(4)         & 5.86(5)           & 0.0053(6)            &                      & 208             & 562                  & 2.1(4)          & C           \\
                &                    &                 &                   &                      &                      & 10              & 759                  & 2.4(3)          & C           \\ \hline
856             & 5/2-, 3/2-, (1/2)- & \textless{}0.1  & \textgreater{}5.0 & 0.0013(9)            &                      & 407             & 449                  & 3.8(5)          & C           \\
                &                    &                 &                   &                      &                      & 208             & 649                  & 0.7(3)          & C           \\
                &                    &                 &                   &                      &                      & 10              & 846                  & \textless 0.7   &             \\ \hline
1001            & 3/2-, (1/2)-       & 1.28(14)        & 5.27(5)           & 0.021(2)             &                      & 407             & 594                  & \textless 0.6   &             \\
                &                    &                 &                   &                      &                      & 208             & 793                  & 5.4(16)         & C           \\
                &                    &                 &                   &                      &                      & 10              & 991                  & 6.7(5)          &             \\
                &                    &                 &                   &                      &                      & 0               & 1001                 & 3.4(4)          &             \\ \hline
1507            & 3/2-, (1/2)-       & 3.20(12)        & 4.73(2)           & 0.070(3)             &                      & 856             & 651                  & 2.1(4)          & C           \\
                &                    &                 &                   &                      &                      & 713             & 795                  & 1.6(6)          & C           \\
                &                    &                 &                   &                      &                      & 698             & 810                  & 1.4(2)          & C           \\
                &                    &                 &                   &                      &                      & 407             & 1100                 & 14.0(8)         & C           \\
                &                    &                 &                   &                      &                      & 208             & 1300                 & 2.2(3)          & C           \\
                &                    &                 &                   &                      &                      & 10              & 1498                 & 17.4(9)         &             \\ \hline
2083            & 3/2-, (1/2)-       & 0.28(3)         & 5.63(5)           & 0.009(1)             &                      & 407             & 1676                 & 3.4(4)          & C           \\
                &                    &                 &                   &                      &                      & 208             & 1875                 & \textless 1.8   &             \\ \hline
2192            & 3/2-, (1/2)-       & 0.55(5)         & 5.31(4)           & 0.019(2)             &                      & 856             & 1335                 & 1.4(2)          & C           \\
                &                    &                 &                   &                      &                      & 407             & 1785                 & 1.8(4)          & C           \\
                &                    &                 &                   &                      &                      & 10              & 2182                 & 3.5(4)          &             \\
                &                    &                 &                   &                      &                      & 0               & 2192                 & \textless 1.3   &      \\ \hline      
\end{tabular}
\end{table*}

\begin{figure*}[ht!]
\resizebox{2.0\columnwidth}{!}{\rotatebox{0}{\includegraphics[clip=]{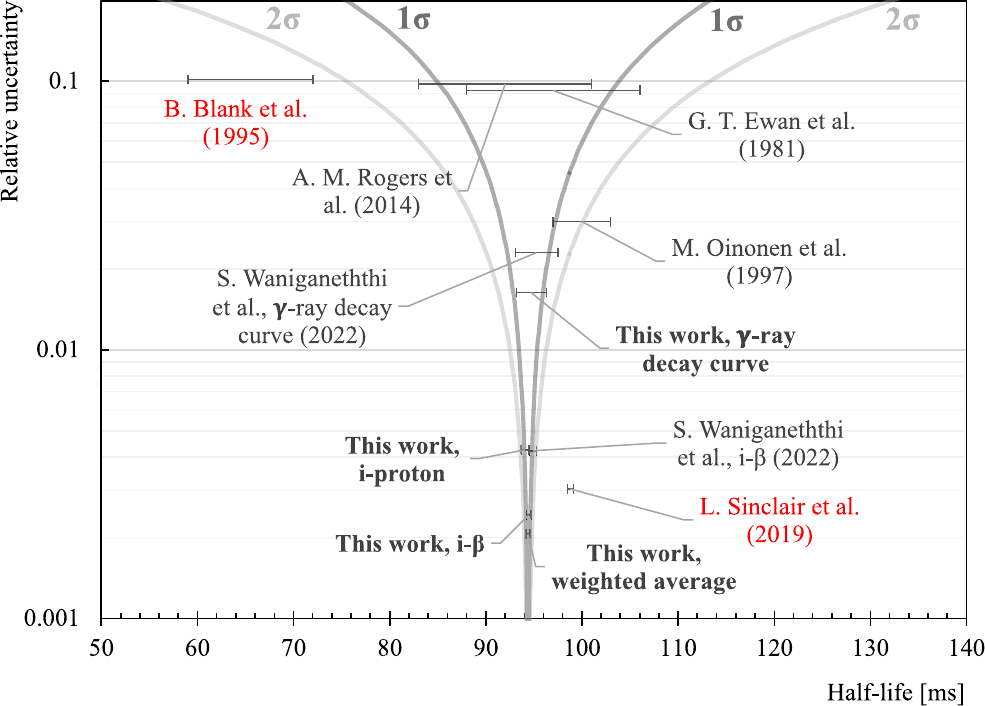}}}
\caption{ 
\label{fig:71Kr_T12}
The present results for the half-life of $^{71}$Kr (the three different methods and their weighted average) is compared with the values from earlier works. The dark gray curve and the light gray curve show the $1\sigma$ and $2\sigma$ confidence intervals for the weighted average of this work, respectively. Except for the outlier results of L. Sinclair {\it et al.} (Phys. Rev. C 100, 044311 (2019)) , where there is probably a typo in the reported uncertainties (private communication of Prof. Wadsworth, Univ. of York) and the value of B. Blank {\it et al.} (Phys. Lett. B 364, 8 (1995)) obtained with limited statistics, all earlier results are consistent within the 2$\sigma$ confidence level.
}
\end{figure*}  

%

\begin{figure*}[htb!]
\resizebox{2\columnwidth}{!}{\rotatebox{0}{\includegraphics[clip=]{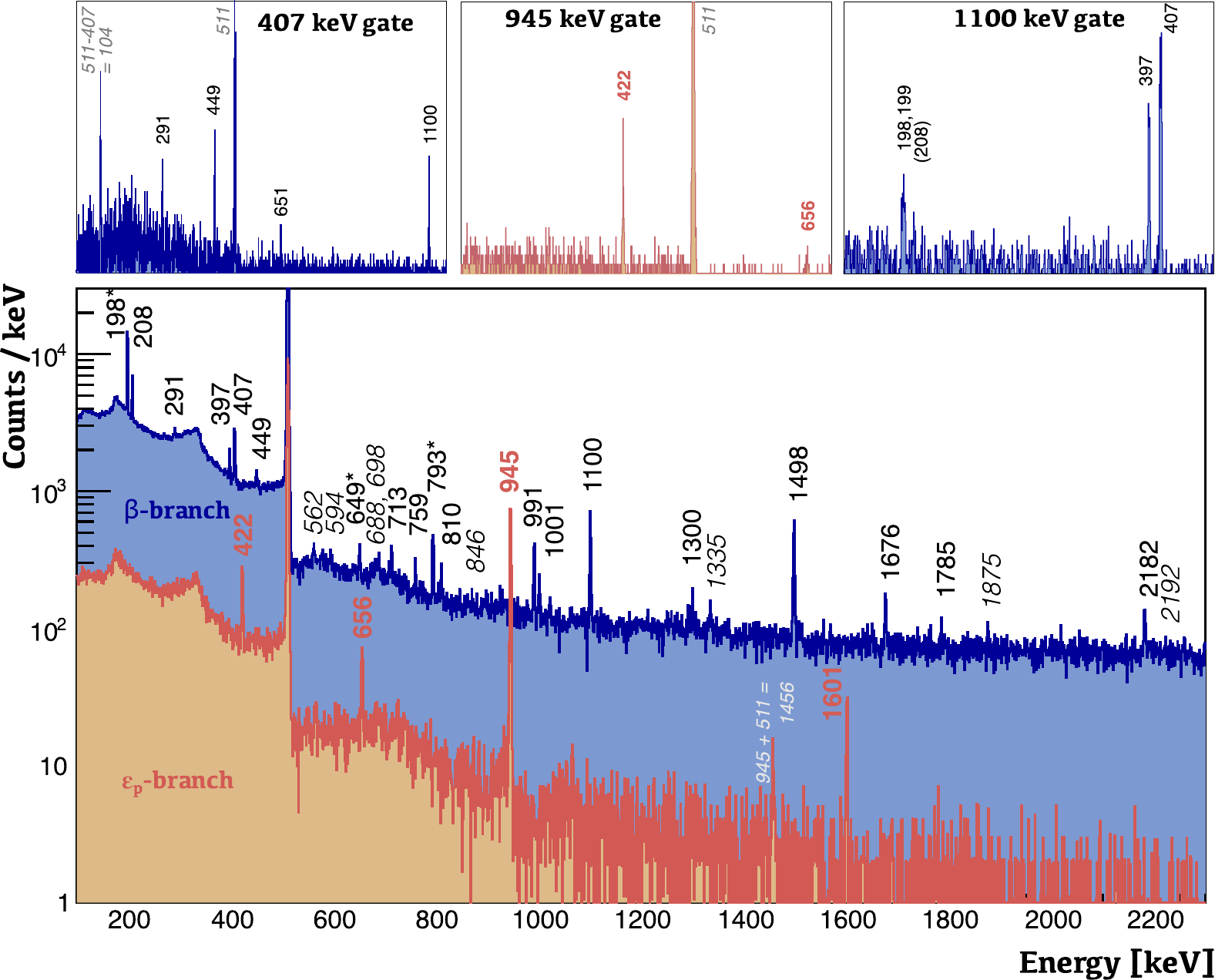}}}
\caption{ 
\label{fig:spectrums}
Energy spectrum of $\beta$-delayed $\gamma$-rays ($\beta$-events with no delayed protons in blue, while orange is used for proton related events). Peak doublets are marked with an asterisk ($^*$). In the upper sub-figures a selection of $\gamma$-gated energy spectra are shown. The 407 keV gated spectrum was used to separate the intensity of the 649/651 keV doublet, while the 1100 keV gated spectrum was used to measure the 199 keV branching ratio of the de-excitation of the 407 keV level. Peaks labeled in grey are related to 511 keV annihilation photons.}
\end{figure*}

\begin{figure*}[ht!]
\resizebox{2.0\columnwidth}{!}{\rotatebox{0}{\includegraphics[clip=]{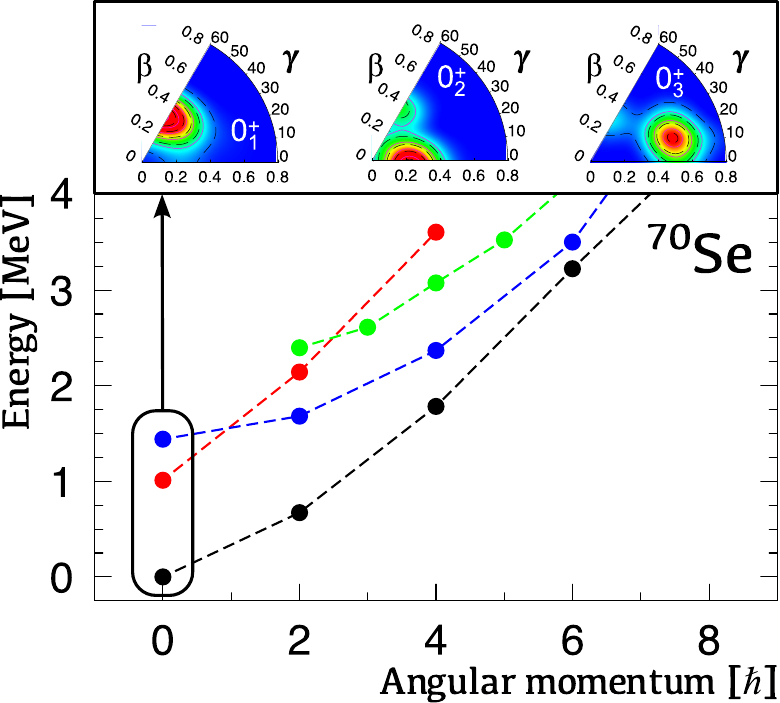}}}
\caption{ 
\label{fig:70Se}
 (top) Collective wave functions for the band-heads in ${70}$Se in the triaxial $(\beta,\gamma)$ plane using SCCM calculations with the Gogny D1S energy density functional. (bottom) Calculated energies for ground state and lowest excited state bands in the same framework.
}
\end{figure*}

\end{document}